\documentclass[11pt]{article}
\usepackage{graphicx,amsmath,amsthm,mathrsfs,amssymb}
\usepackage[usenames]{color}
\usepackage{ulem}

\newcommand{\ee}{\end{equation}}

\newcommand{\reff}[1]{(\ref{#1})}
\newcommand{\beq}{\begin{equation}}
\newcommand{\eeq}[1]{\label{#1}\end{equation}}
\newcommand{\beqa}{\begin{eqnarray}}
\newcommand{\eea}{\end{eqnarray}}
\newcommand{\eeqa}[1]{\label{#1}\end{eqnarray}}
\newcommand{\beg}{\begin{equation*}}
\newcommand{\eeg}{\end{equation*}}

\newcommand{\eq}{\!=\!}

\newcommand{\bsplit}{\begin{split}}
\newcommand{\esplit}{\end{split}}

\begin{document}
\begin{titlepage}
\title{Directly observing entropy accumulate\\ on the horizon and holography}
\author{\thanks{aedery@ubishops.ca} Ariel Edery and \thanks{hbeauchesne10@ubishops.ca} Hugues Beauchesne \\\\{\small\it Physics Department, Bishop's University}\\
{\small\it 2600 College Street, Sherbrooke, Qu\'{e}bec, Canada
J1M~0C8}}
\date{March 31, 2012.}
\maketitle
Essay written for the Gravity Research Foundation 2012 Awards for Essays on Gravitation
\begin{abstract}
Recent numerical simulations of gravitational collapse show that there exists a special foliation of the spacetime where matter and entropy accumulate directly on the inside of the horizon surface. In this foliation, the time coincides with the proper time of the asymptotic static observer (ASO) and for spherical symmetry, this corresponds to isotropic coordinates. In this gauge, the three-volume in the interior shrinks to zero and only the horizon area remains at the end of collapse. In a different foliation, matter and entropy accumulate in the volume. The entropy is however independent of the foliation. Black hole holography is therefore a mapping from an arbitrary foliation, where information resides in the volume, to the special ASO frame, where it resides directly on the horizon surface.        
   \end{abstract}
\end{titlepage}
In the early 1970's Jacob Bekenstein discovered that black holes have an entropy proportional to the area of the event horizon and Stephen Hawking later fixed the proportionality constant to one quarter (in Planck units). This does not imply that information cannot reside in the interior volume but that it scales with horizon area instead of volume. This is an example of the holographic principle \cite{tHooft,Susskind} since three-dimensional information is mapped onto or encoded in a two-dimensional boundary. This raises an important question: can one view directly, explicitly, the two-dimensional nature of the entropy in a dynamical setting? The answer is yes. Recent numerical simulations of spherically symmetric gravitational collapse \cite{Khlebnikov, C-E, B-E, B2-E} have shown that there exists a foliation where the entropy and matter can be observed to accumulate directly on the surface of the horizon. In this foliation, the time coincides with the proper time of an asymptotic static observer (ASO). For spherical symmetry, this corresponds to isotropic coordinates. In this gauge, the interior three-volume shrinks to zero by the end of the collapse. In other foliations, matter and entropy accumulate in the interior volume not on the horizon surface. But different foliations must yield the same entropy. Black hole holography is therefore a mapping from an arbitrary foliation, where the information resides in the volume, to the foliation corresponding to the ASO frame, where the information resides directly on the horizon. We will return to this point later. For now, we will see how matter can accumulate on the horizon instead of the volume during gravitational collapse. This is very important, because until recently, most numerical simulations of gravitational collapse did not observe this phenomena.
    
Consider a test-particle falling radially towards a Schwarzschild black hole. Freely falling observers who follow the particle will see it cross the horizon in a finite time; the matter does not accumulate at the horizon. The proper time of freely falling observers corresponds to the coordinate time in Painlev\'e-Gullstrand (PG) coordinates and numerical simulations of gravitational collapse in this gauge show that the matter accumulates in the interior volume at $r\eq 0$ \cite{Kunstatter}. In contrast, from the viewpoint of an ASO, it is well known that the particle takes an infinite time to reach the horizon; it approaches the horizon but never crosses it. The proper time measured by the ASO is the coordinate time in standard Schwarzschild coordinates and we will refer to it as ``Schwarzschild time". As more matter is thrown into the black hole, it becomes squeezed into a thin shell just outside the horizon as the Schwarzschild time increases (the ``thin pancakes" at the horizon discussed in \cite{Lindesay}). However, there is a serious obstacle to overcome before this foliation, the ASO frame, can be adopted in a dynamical collapse setting. Standard Schwarzschild coordinates cannot be used for numerical studies of collapse because there is a coordinate singularity at the horizon. This is typically resolved by introducing new coordinates that are a combination of the Schwarzschild time and the radial coordinate; you can ``fix" the coordinate singularity but at the cost of abandoning Schwarzschild time. The common gauges used for numerical studies of gravitational collapse such as maximal slicing, PG, Eddington-Finkelstein, double-null, etc. do not have a coordinate corresponding to Schwarzschild time. This is precisely the reason why, until recently, numerical simulations did not observe a thin shell of matter forming near the horizon during collapse to a black hole.    

\begin{figure}[tbp]
		\begin{center}
			\includegraphics[scale=.38, draft=false, trim=1.5cm 1.5cm 2.5cm 2cm, clip=true]{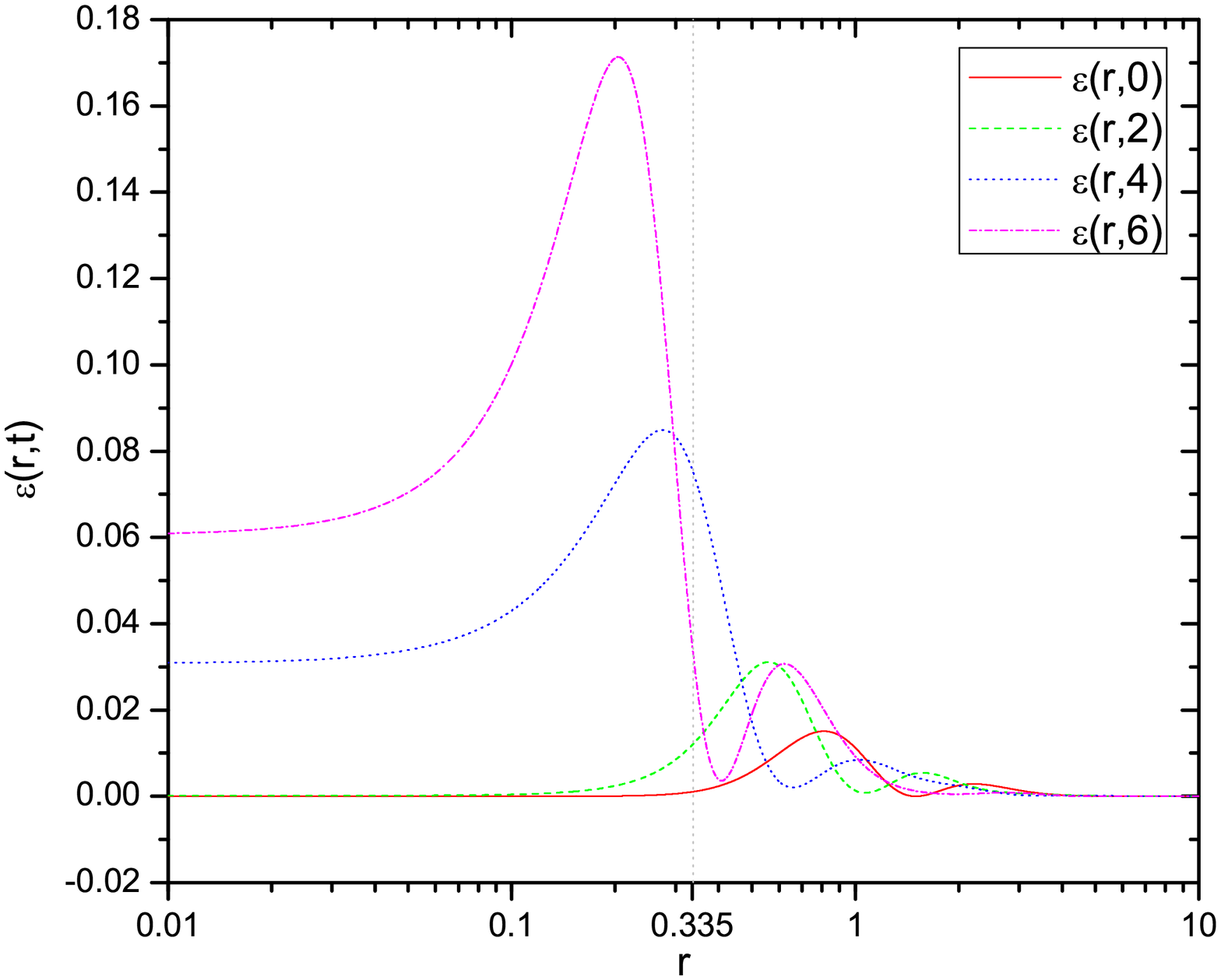}
		\end{center}
		\caption{\label{rho1} The energy density $\epsilon$(r,t) at early times. There are two peaks instead of one because the scalar field has a Gaussian-type distribution (i.e. the derivative of a Gaussian has two peaks). The horizon is at $r_0=0.335$ where $N=0$ at late times. At $t=0$, both energy density peaks are in the interior ($r<r_0$). Both peaks evolve towards the left, towards smaller values of $r$, increase in height and become thinner (their width becomes smaller). They enter into the interior region (see next graph Fig.\reff{rho2} for late time evolution).}
\end{figure}

\begin{figure}[tbp]
		\begin{center}
			\includegraphics[scale=.38, draft=false, trim=1.5cm 1.5cm 2.5cm 2cm, clip=true]{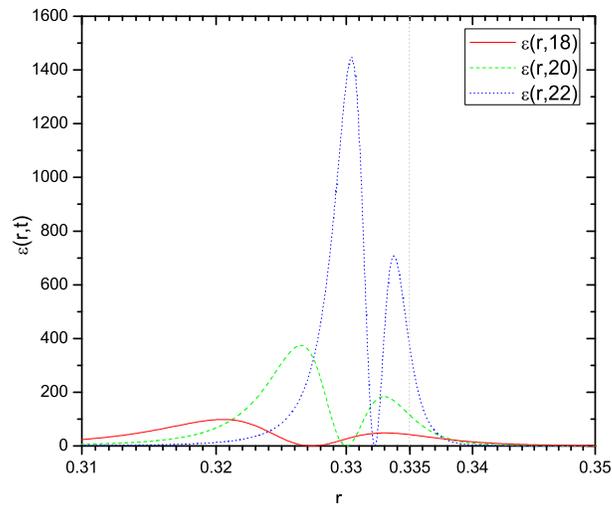}
		\end{center}
		\caption{\label{rho2} The energy density at late times. Both peaks are in the interior at $t=18$ and are close to the horizon at $r=0.335$. Note that the two peaks move now radially towards the {\it right}, towards the horizon as time evolves. The energy density is getting squeezed into a thin shell just inside the horizon (the peaks are found in the region $0.335>r>0.33$ at $t=22$, the latest available time). In the exterior, the energy density extends out less with time. }
\end{figure}

\begin{figure}[tbp]
		\begin{center}
			\includegraphics[scale=.38, draft=false, trim=1.5cm 1.5cm 2.5cm 2cm, clip=true]{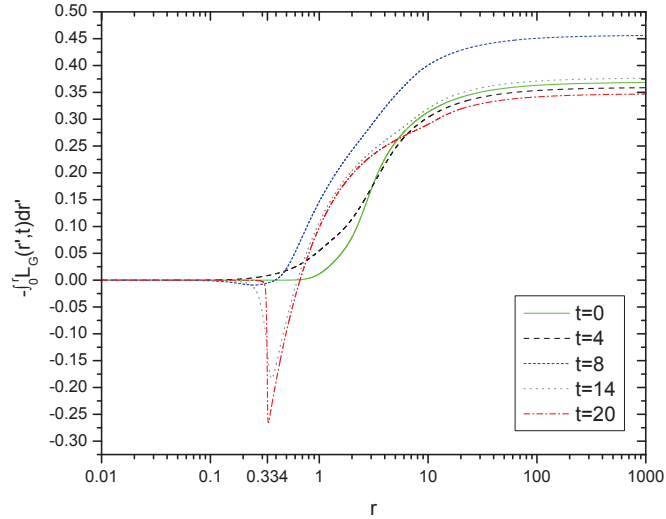}
		\end{center}
		\caption{\label{Lang1} The negative of the Lagrangian, corresponding to the free energy, plotted at different times during 4D collapse of a massless scalar field \cite{C-E}. Note the sharp negative dip inside and near the horizon. The entropy accumulates in this thin region.}
		\end{figure}

It turns out that one can fix the coordinate singularity and maintain Schwarzschild time. The ``trick" is to write the spatial part of the metric in a form which is conformal to flat and {\it not alter the definition of time}. This leads to the isotropic coordinate system \cite{Finelli1,Finelli2}. A metric which is spherically symmetric expressed in isotropic coordinates takes the form 

\beq
ds^2= - N^2(r,t)\, dt^2 +\psi^4(r,t) (dr^2 +r^2 d\Omega^2)
\eeq{iso}
where $N$ is called the lapse, $\psi$ the conformal factor and $r$ is the isotropic radial coordinate (which is distinct from the areal radius $R=\psi^2 \,r$). The time $t$ in \reff{iso} is the proper time measured by an ASO.  Numerical simulations in isotropic coordinates have been used to study black hole thermodynamics for the collapse of a 5D Yang-Mills instanton \cite{Khlebnikov}, of a 4D and 5D massless scalar field \cite{C-E, B-E} and recently for 4D charged matter \cite{B2-E}. For the 4D Schwarzschild black hole, the analytical solution in the static exterior is $\psi\eq 1+ M/2r$ and $N\eq (1-M/2r)/(1+M/2r)$ where $M$ is the black hole mass. The horizon in isotropic coordinates occurs at $r_0=M/2$ where $N=0$. Note that $\psi$ does not diverge at the horizon; it is finite and equal to $2$. There is no coordinate singularity. During gravitational collapse, we expect $N$ and $\psi$ in the exterior region ($r\ge M/2$) to approach their static analytical values. However, in the interior ($r<M/2$), they are time-dependent and have been probed only numerically. In isotropic coordinates we observe matter accumulate in a thin shell near the horizon at late times of collapse. However, there is a crucial difference with the test-particle scenario we previously discussed. There is a thin shell of matter at the horizon but it resides mostly on the {\it inside} of the horizon not on the outside. This can be seen clearly in figures \reff{rho1} and \reff{rho2} that show the early and late time evolution respectively of the energy density during the collapse of a massless scalar field \cite{B-E}. At the $N\eq 0$ surface, radial null geodesics have $dr/dt\eq 0$, which explains why matter accumulates at the horizon. However, the radius $r_0$ where $N\eq 0$ increases during the dynamical collapse process and during this time matter in the exterior enters into the interior (see \cite{B-E} for details). At late times, the $N\eq 0$ surface is basically stationary at $r_0\eq 0.335$ and the rate of matter entering in slows to a crawl. By this time, almost all of the matter has entered into the interior and has been squeezed into a thin shell on the inside of the horizon (the situation at $t=22$ in Fig.\reff{rho2}). The entropy of the black hole also accumulates in this thin shell region. We know this from tracking the Lagrangian. There is now considerable numerical evidence from thermodynamic studies of collapse \cite{Khlebnikov, C-E, B2-E} that the negative of the Lagrangian is equal to the Helmholtz free energy $F\eq E-TS$ of a Schwarzschild black hole ($E$ is the black hole mass, $T$ the temperature and $S$ the entropy). It can also be shown analytically that the interior contribution to the free energy is negative, stems from the dynamical region \cite{Vaz1,Vaz2,E-C} and is equal to $-T\,S$ \cite{Khlebnikov, C-E}. Numerically, this contribution appears near the horizon. This can be seen in Fig.\reff{Lang1} where the negative of the Lagrangian for 4D collapse of a massless scalar field \cite{C-E} is plotted. The entropy clearly accumulates in a thin slice just inside the horizon. 

\begin{figure}[tbp]
		\begin{center}
			\includegraphics[scale=.38, draft=false, trim=1.5cm 1.5cm 2.5cm 2cm, clip=true]{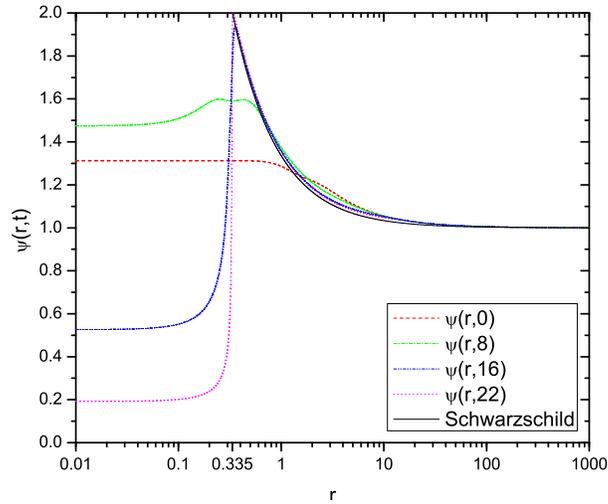}
		\end{center}
		\caption{\label{psi1} The conformal factor $\psi$ at different times. In the interior, $\psi$ is positive and is decreasing towards zero. As one approaches the horizon ($r=0.335$) from the interior, $\psi$ increases sharply towards a value of $2$, its analytical expectation. The exterior matches closely the Schwarzschild analytical form.}
		\end{figure}

\begin{figure}[tbp]
		\begin{center}
			\includegraphics[scale=.38, draft=false, trim=1.5cm 1.5cm 2.5cm 2cm, clip=true]{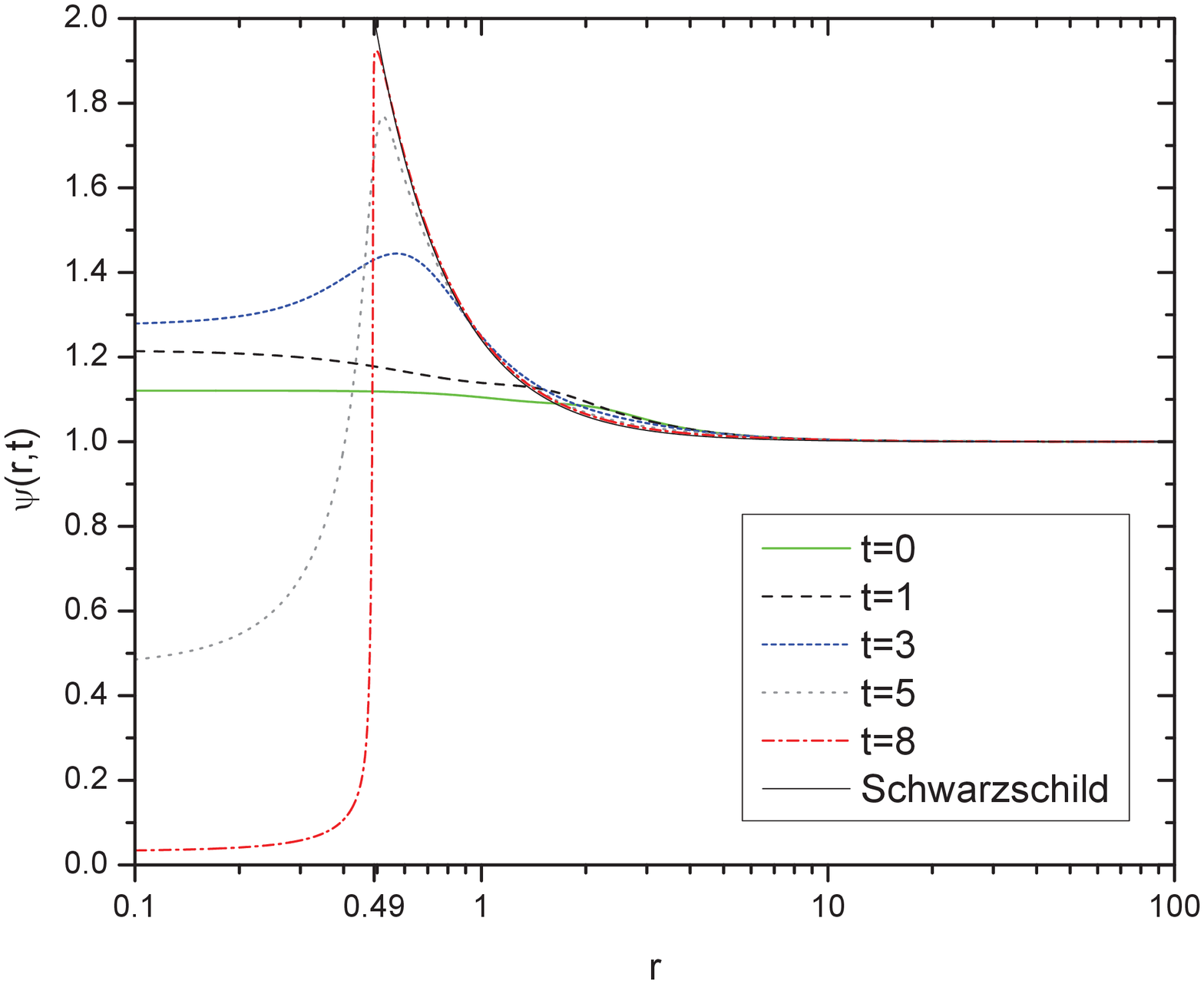}
		\end{center}
		\caption{\label{psi2} The conformal factor $\psi$ in five-dimensional collapse. Again, in the interior $\psi$ is decreasing towards zero and matches the Schwarzschild analytical form in the exterior. It has a value of $2$ near the horizon at $r=0.49$.}
		\end{figure}

In Fig.\reff{psi1}, the metric function $\psi$ is plotted  \cite{B-E}. In the interior, $r<r_0$, $\psi$ is homogeneous and is decreasing towards zero during the collapse (this also occurs in 5D, see Fig.\reff{psi2}). The proper radial distance from $r\eq0$ to the horizon at $r_0$, $\int_0^{r_0} \psi^2 dr \approx \psi^2\,r_0$, approaches zero during the collapse. The radial location of the thin shell of matter remains at the horizon $r_0$ while this occurs. In the interior, as $\psi\to 0$, the three-volume of a spacelike hypersurface at a constant time slice ($4\pi\int_0^{r_0} \psi^6\,r^2\,dr \approx \psi^6\, 4\pi\,r_0^3/3$) clearly tends to zero. There is no volume in the interior at the end of the collapse. As one approaches the horizon radius $r_0$ from the inside, $\psi$ rises sharply to the integer value of $2$. In other words, within a tiny slice inside the horizon, $\psi$ is not zero and the areal radius at the horizon is $R\eq \psi^2\,r_0 \approx 4\,r_0 =2\,M$ where $M\equiv 2\,r_0$ is the black hole mass. At the end of the collapse, all that remains of the interior is the horizon area $A\eq 16\,\pi\,M^2$.  

In other foliations, where coordinate time is not Schwarzschild time, the radial velocity $dr/dt$ of null geodesics is not zero on the horizon and this is why matter and information do not accumulate there but in the interior volume. For example, we have already seen that in PG coordinates the matter accumulates in the interior at $r=0$. Black hole entropy is an invariant; the amount of information hidden behind the event horizon does not depend on the foliation. Black hole holography is therefore a mapping from an arbitrary foliation, where information resides in the volume, to the special ASO frame, where it resides directly on the horizon. 

The interior three-volume of a black hole is not an invariant. This is precisely why black hole entropy cannot scale with volume. The three-volume depends crucially on the foliation of the spacetime. For example, in isotropic coordinates it is zero and in PG coordinates, because its spatial sections are flat \cite{Poisson}, it is $4\pi R^3/3 $ with $R=2M$.  In contrast, the area of the event horizon is an invariant \cite{Wald} and this is why it can act as the entropy. However, the two-dimensional nature of black hole entropy is {\it explicit} only in a particular foliation.

\newpage	         

\begin{thebibliography}{99}
\bibitem{tHooft} G. 't Hooft, [arXiv:gr-qc/9310026].
\bibitem{Susskind} L. Susskind, J. Math. Phys. {\bf 36}, 6377 (1995) [arXiv:hep-th/9409089].
\bibitem{Khlebnikov} Z. Gecse and S. Khlebnikov, Phys. Rev. D {\bf 77}, 104003 (2008) [arXiv:0801.3662].
\bibitem{C-E} B. Constantineau and A. Edery, Phys. Rev. D {\bf 84}, 084032 (2011) [arXiv:1103.5272].
\bibitem{B-E} H. Beauchesne and A. Edery, Phys. Rev. D {\bf 85}, 044056 (2012) [arXiv:1108.0449]. 
\bibitem{B2-E} H. Beauchesne and A. Edery, [arXiv:1203.2279].
\bibitem{Kunstatter} J. Ziprick and G. Kunstatter, Phys. Rev. D {\bf 79}, 101503 (2009) [arXiv:0812.0993].
\bibitem{Lindesay} L. Susskind and J. Lindesay, {\it An introduction to black holes, information and the string theory revolution: the holographic universe} (World Scientific Publishing, London, 2005).
\bibitem{Finelli1} F. Finelli, and S. Khlebnikov, Phys. Lett. B {\bf 504}, 309 (2001).
\bibitem{Finelli2} F. Finelli and S. Khlebnikov, Phys. Rev. D {\bf 65}, 043505 (2002).
\bibitem{Vaz1} C. Vaz, Phys. Rev. D {\bf 61}, 064017 (2000).
\bibitem{Vaz2} C. Vaz and L. Witten, Phys. Rev. D {\bf 64}, 084005 (2001). 
\bibitem{E-C} A. Edery and B. Constantineau, Class. Quantum Grav. {\bf 28}, 045003 (2011) 
[arXiv:1010.5844]
\bibitem{Poisson} E. Poisson, {\it A Relativist's Toolkit} (Cambridge University Press, Cambridge, 2004).
\bibitem{Wald} R. Wald, General Relativity (University of Chicago Press, Chicago, 1984).
\end{thebibliography}
\end{document}